\documentstyle[12pt,aasms4]{article}
\received{----------- }
\accepted{----------- }
\journalid{337}{ }
\articleid{11}{ }
\slugcomment{}
\lefthead{Christine C. Dantas, Andr\'e L. B. Ribeiro, Hugo V. Capelato \& Reinaldo R. de Carvalho}
\righthead{The Two-component Virial Theorem and the Physical Properties
of Stellar Systems}

\begin{document}

\title{THE TWO-COMPONENT VIRIAL THEOREM AND\\
THE PHYSICAL PROPERTIES OF STELLAR SYSTEMS}

\author{Christine C. Dantas\altaffilmark{1}}
\author{Andr\'e L. B. Ribeiro\altaffilmark{2}}
\author{Hugo V. Capelato\altaffilmark{1}}
\author{Reinaldo R. de Carvalho\altaffilmark{3}}

\altaffiltext{1}{Divis\~ao de Astrof\'{\i}sica, INPE/MCT, CP 515, S. J. dos Campos, 12201-970 SP, Brazil.}
\altaffiltext{2}{Departamento de Matem\'atica
Aplicada, IMECC, Universidade Estadual de Campinas,
13083-970 SP, Brazil. }
\altaffiltext{3}{Observat\'orio Nacional, Rua Gal.  Jos\'e Cristino,
77 -- 20921-400, Rio de Janeiro, RJ., Brazil}

\begin{abstract}

Motivated by present indirect evidences that galaxies are surrounded by
dark matter halos, we investigate whether their physical properties can
be described by a formulation of the virial theorem which explicitly
takes into account the gravitational potential term representing the
interaction of the dark halo with the barionic or luminous component.
Our analysis shows that the application of such a ``two-component
virial theorem'' not only accounts for the scaling relations displayed,
in particular, by elliptical galaxies, but also for the observed
properties of all virialized stellar systems, ranging from globular
clusters to galaxy clusters.

\end{abstract}


\section{Introduction}

It is expected on very fundamental grounds that the state of
equilibrium of self-gravitating, time-averaged stationary stellar
systems should be well described by the virial theorem.  In fact,
elliptical galaxies, for instance, show a remarkable homogeneity,
expressed by a very tight kinematical-structural relationship, the
so-called ``Fundamental Plane'' (FP, \cite{djo87}, \cite{dre87}).
Since it is believed that these galaxies represent equilibrium systems,
their interconnected physical properties should reflect their
virialized condition. However, the FP is significantly ``tilted''
relatively to the relations expressed by the virial theorem applied to
a family of homologous objects.  The nature of this discrepancy is
controversial and has been extensively debated in the literature (e.g.
\cite{gc97}, Ciotti, Lanzoni \& Renzini 1996, \cite{pah98} and
references therein).

The FP ``problem'' can be stated as follows. The virial theorem, applied
to a stationary self-gravitating system states that  $2K + W = 0$,
where $K$ is the kinetic energy and $W$ is the potential energy of the
system.  This may be re-written as $ \langle v^2\rangle = {GM/r_G}$,
where $r_G$ is the gravitational radius, defined by $r_G = GM^2/|W|$,
$\langle v^2 \rangle$ is the mean square velocity of the particles, $G$
is the gravitational constant, and $M$ is the total mass of the
system.  These physical quantities may be translated to observational
ones through the definition of some kinematical-structural coefficients
($C_r, C_v$) which may or may not be constants among galaxies:
$\sigma_0^2 = C_v \langle v^2 \rangle $ and $r_e = C_r r_G$ ; $I_e =
\left ( {M/2 \over \pi r_e^2} \right ) \left ({M \over L} \right )
^{-1}$. $M / L$ is the mass-luminosity relation for the system; $r_e$
its {\it effective radius},  that is the radius which contains half of
its total luminosity:  $L(<r_e) = L_{tot}/2$; $\sigma_0$ its {\it
central projected velocity dispersion}, that is the mean square
projected velocity of stars at the galaxy center (measured inside a
slit of finite projected width); and  $I_e = {L(<r_e)/\pi r_e^2}$, is
the {\it mean surface brightness inside} $r_e$ in linear units.
Inserting these equations into the virial relation one finds: $ r_e =
C_{fp} \sigma_0^2 I_e^{-1}$, where $C_{fp}$ depends on the
mass-luminosity relation and on the coefficients defined above ($C_r,
C_v$). In contrast, what one observes is that $ r_e \propto \sigma_0^A
I_e^B$, with $A \sim 1.53$, $B \sim -0.79$, for elliptical galaxies
observed in the near-infrared (\cite{pah98}). The reasons for the
deviation of the observed relationship as compared to the virial
theorem are not well established. One may postulate a systematic
variation of the structural coefficients (galaxies would form a
non-homologous family of objects:  \cite{CdCC95}; \cite{hm95}), or
yet a systematic trend of the mass to light ratio with galaxy mass: ${M
/ L} \propto M^{\alpha}$ (e.g. \cite{dre87}).

However, it should be noted that elliptical galaxies, as any other
collapsed structures, are probably surrounded by massive dark matter
halos. The observed FP relations, on the other hand, arise from the
observed (i.e., {\it barionic}) component of these systems. It seems
thus natural to ask how the equilibrium state of the barionic component
{\it under} the influence of its massive halo would modify the simple
one component virial theorem. In fact, attempts to construct
two-component models can be found in the recent literature. For
instance, \cite{clr96} indicate that the FP tilt could be explained by
massive extended dark matter halos embedding the luminous matter of
galaxies with the following caveat:  a non-realistic fine-tuning of the
luminous-to-dark matter distributions  would be required in order to
explain the small scatter of the FP correlations.  On the other hand,
preliminary results of \cite{k97} suggest that the FP for ellipticals
and, also, the observed deviation of dwarf spheroidal galaxies from
it, may follow from the dynamical equilibrium condition in the
framework of a two-component model.

Attempting to visualize the physical properties of virialized stellar
systems of various scales into an integrated framework (the
$\kappa$-space, c.f. \cite{ben92}), \cite{bur97} (hereafter BBFN97)
concluded that globular clusters, galaxies, groups of galaxies and
clusters of galaxies also show systematic trends in their observed
properties, populating what they called a ``cosmic metaplane'' in their
parameter space. This metaplane, also tilted {\sl wrt} the simple
virial expectation, was interpreted as a combination of FP-like tilts
associated to the various stellar systems, possibly reflecting their
different stellar population and dissipation histories. However, under
this interpretation, a fine-tuning mechanism for the variation of $M/L$
with mass, for every stellar system, had also to be invoked in order to
preserve the striking appearance of the metaplane (see also
\cite{sch93}). Also, their analysis made evident a ``zone of
exclusion'' (ZOE) where no stellar system could be found.  This raises
the question of which formation process would generate such a trend and
the mechanisms responsible for producing the metaplane itself.

In this {\sl Letter} we tackle these questions by starting from the
hypothesis that self-gravitating stellar systems in the universe are
embedded in dark halos. As a consequence, the strict virial theorem
must be replaced by a new equilibrium equation which takes explicitly
into account the gravitational potential produced by  the massive halo in
which is embedded the luminous component. With this assumption, we
present an alternative model which may naturally explain the issues
discussed above.  Our paper is organized as follows: in
Section 2, we discuss the virial theorem for two-component systems; in
Section 3, we apply it to observational data; and in Section 4, we
discuss some of the implications of our results.

\section{The Two-Component Virial Theorem}

The scalar virial theorem for the barionic component of a stellar
system (component-2), in steady-state equilibrium embedded in its dark
matter halo (component-1), may be readily deduced from the Jeans
equation by assuming that, in addition to its self-potential, it is
also subjected to the external potential produced by the dark matter
(see e.g. \cite{bet87}; see also \cite{limb59}, \cite{spi69}, \cite{smith}). In
this case a new term is added to the gravitational energy of the system
due to the interaction of the two components. Assuming spherical
symmetry we may write the gravitational energy of luminous component,
$W_2$, as:
$$ W_2 =
- G \int_0^{\infty} {\rho_2(r) M_2(r) \over r} dV - G \int_0^{\infty}
{\rho_2(r) M_1(r) \over r} dV
\eqno(1)$$
where $M_{\mu}(r)$ is the total mass of the $\mu$-component, within the
radius $r$. If we now further assume that the dark matter halo
~-~component-1~-~ is more extended than the barionic component, having
a not too steep density profile within the interior region containing
the luminous component, then we may approximate the second integral,
which gives the interaction energy, by:
$$ W_{21} \equiv - G \int_0^{\infty} {\rho_2(r) M_1(r)
\over r} dV \sim -{4 \pi \over 3} \rho_{0,1} G \int_0^{\infty} {\rho_2(r) r^3
\over r} dV = -{4 \pi \over 3} \rho_{0,1} G M_2 \langle r_2^2 \rangle
\eqno(2)$$
where $\rho_{0,1}$ is the mean density of the dark matter halo within
the region containing the luminous component and
$$
\langle r_2^2 \rangle \equiv {\int r^2 \rho_2(r) dV \over \int \rho_2(r) dV}
\eqno(3)$$
Thus, the virial theorem for the collapsed barionic component, $2K_2 +
W_2 = 0$, may be written as:
$$
\langle v_2^2 \rangle = {G M_2 \over r_{G,2}} + {4 \pi \over 3} G \rho_{0,1}
\langle r^2_2 \rangle
\eqno(4)$$
where $r_{G,2}$ is  the gravitational radius of the second component.

We see that in the presence of an extended dark matter halo the virial
theorem gets an extra term on its right hand side, which accounts for
the interaction with the extended dark matter halo (this is also known
as the ``Limber effect''). As we will see in the next section, this
term is essential for our understanding of the systematic trends of the
observed properties of the stellar systems we discussed before.  In
terms of the observational quantities the modified virial theorem
writes as:
$$
\sigma_0^2 = C^* (I_e r_e + b r_e^2) ~~{\rm where}~~
C^* = 2\pi G C_r C_v \left ( {M\over L} \right )_2
\eqno(5)$$
and
$$b = {2 \over 3} {R\over C_r} \left ( {M\over L} \right )^{-1}_2
\rho_{0,1}~~~~{\rm with}~~~~R = {\langle r_2^2 \rangle \over r_e^2}
\eqno(6)$$

Notice that in these equations all the structural coefficients as well
as $M/L$ refer to the barionic component.  Parameter $b$ has dimension
of a luminosity density whereas $C^*$ has dimension of a less intuitive
quantity (i.e., $GM/L$).

Eq.(6) is specially interesting, since it relates the parameter $b$ to
the central density of the dark matter halo. We numerically analyzed
various equilibrium models (specifically, Jaffe, King and Sersic
models, c.f.  \cite{bet87}, \cite{c91}, \cite{cl96}) and found that
$C_r C_v \sim 0.2$, whereas $R/C_r$ varies significantly, depending on
the models: $R/C_r$ $\sim 10 - 25$ for King or Jaffe models and $\sim
10 - 60$ for the Sersic models. We adopted $R/C_r$ $\sim 20$ as a
typical value. It is important to stress that for galaxies this
approximation can introduce a factor of 2 difference in the parameter
b.

\section{Applying the Two-Component Virial Theorem }

We will apply the two-component virial theorem (2-VT) in the context of
the $\kappa$-space parameter framework. This will allow us to directly
compare the 2-VT predictions with the extensive data provided by
BBFN97. In this coordinate system the FP is seen edge-on, projected on
the ($\kappa_1 , \kappa_3$) plane, and the two-component virial theorem
(Eq. (5)) can be expressed as :
$$
\kappa_3 = {\log C^* \over \sqrt{3}} + {1 \over \sqrt{3}} \log
\left ( 1 + b 10^{\omega} \right )
\eqno(7)$$
where
$$\omega \equiv (\kappa_1 - \sqrt{3} \kappa_2)/\sqrt{2} = -\log I_e/r_e
\eqno(8)$$
that is, $\omega$ is measuring the central luminosity density of the
stellar systems.

>From Eqs. (7) and (8) we see that the 2-VT defines a surface in the
$\kappa$-space which main characteristics may be best viewed through
the curve defined by its intersection with the ($\kappa_3 , \omega$)
plane, perpendicular to the ($\kappa_1 , \kappa_2$) plane. A brief
analysis of Eq. (7) shows that it intercepts the $\kappa_3$-axis
 at ${\log C^* /\sqrt{3}}$. If there were no dark halo, $b = 0$,
recovering the usual 1-component virial theorem, $\kappa_3 = cte$.
Departure from this horizontal line at a given $\omega$ depends on the
term $ b 10^{\omega}$ and thus on the density of the dark halo.  For $b
10^{\omega} >> 1$ it tends to a straight line with a fixed slope of ${1
/ \sqrt{6}}$ intercepting the $\kappa_3$-axis at ${\log(C^*b) /
\sqrt{3}}$.  That is, the 2-VT predicts an {\it asymptotic,
characteristic, fixed tilt} relatively to the 1-component virial
theorem. Notice that, within a factor depending on the structural
coefficients of the barionic component, the value of the mean central
density of the dark matter halo is given by this intercept.

In Figure 1a, we plot the data on the $\kappa$-space, projected on
($\kappa_1 , \kappa_3$) plane for self-gravitating stellar systems
spanning all scales, from globular clusters to rich galaxy clusters,
using data presented in BBFN97. The two-component virial theorem
curves, given by $C^* = 8.28$ and $b = 200$, are shown in dotted-line
for the various ranges of the $\kappa$ parameters.  This figure shows
the striking compatibility of the ``cosmic metaplane'' with the
theoretical predictions of the two-component virial theorem -
specifically, {\it the fixed asymptotic tilt relatively to the strict
virial theorem}.

We establish the 2-VT relation (Eq. (7)) by assuming two different
hypothesis about the mass-luminosity relation of the barionic
component: $(a)$ that its value is about the same as found for the
globular clusters, which seems reasonable since these systems are very
well described by the 1-component virial theorem, that is $b
10^{\omega_{glob~clust}} << 1$ (see \cite{bel98}); and $(b)$ by
adjusting the value of the $\kappa_3$ intercept (that is, $(M/L)_2$),
to a maximum value still giving a reasonable fit to the groups and
clusters of galaxies. In doing that we were attempting to take into
account the presence non-stellar barionic mass and also for remaining
galactic dark halos in these systems.

We find for case $(a)$ $(M/L)_2 \sim 1.6$ ($C^* = 8.28$), a value which
agrees fairly well with those for globular clusters (e.g.
\cite{pry93}). For case $(b)$, $C^* = 39.2$, gives $(M/L)_2 \sim 7.4$.
The central densities of the dark matter halos were estimated after
adjusting the $b$ parameter. For the galaxies (ellipticals) we found
$\rho_{0,1} \sim 2.3 \times10^{-2}~ M_{\odot}/pc^3$ ($b = 200$),
whereas for the elliptical dominated groups and clusters of galaxies,
$\rho_{0,1} \sim 5.8  \times10^{-6}~ M_{\odot}/pc^3$ ($b = 0.20$, for
case $(a)$ and $b = 0.004$, for case $(b)$). The corresponding values
for the spiral galaxies and for the spiral dominated groups are about a
factor 2 -- 3 smaller due to the fact that these systems appear
slightly displaced towards larger values of $\omega$.

In Figure 1b we clearly see that the points do not fill the space
continuously. On the contrary, they are arranged in some bands defined
by specific values of $\kappa_3$, which are related to specific values
of $w$ through Eq.(7). The parameter $w$ rules the luminosity density
in the systems and hence it is associated with their dissipation
histories and the epoch when the collapse happened (i.e. the density
fluctuation spectrum). Thus, in the context of a hierarchical
clustering scenario, smaller systems collapse before and are more
concentrated, presenting higher luminosity densities ($w$ more
negative); while larger objects, collapsed later, present lower
luminosity densities ($w$ more positive). The scatter in the
perpendicular direction to $w$ probably reflects a change in mass which
produces the bands seen in Figure 1b. The gaps between different
objects on the ($\kappa_1 , \kappa_2$) plane were firstly noted by BBFN97,
but now we have  quantified this feature by the parameter $w$. A full
account of the role of the parameter $w$ is beyond the scope of the
present Letter.

\section {Discussion}

This work is based on  the hypothesis that self-gravitating, equilibrium
stellar systems in general possess an extended dark matter halo.  In
order to describe their equilibrium state, a modified, two-component
virial theorem must be taken into account which predicts the existence
of a fundamental surface. We found a remarkable compatibility of this
hypothesis with the observed properties of a great range of stellar
systems. Particularly, the ``cosmic metaplane'', first discussed by
BBFN97 as an  ensemble of interrelated fundamental planes, is shown to
reasonably follow the fundamental surface here derived.

Furthermore our analysis reinforces the view that the FP relations
should arise as a correction to the observed (luminous) parameters
relations for the presence of the dark (unseen) matter surrounding
these systems. However, as pointed out by \cite{clr96}, this does not
completely solve the FP problem, since a fine-tuning of the
dark-to-luminous matter distributions is required in order to explain
the small scatter of the FP correlations. Although no such a mechanism
has been proposed or known up to now, there is at least one piece of
evidence that it may exist, as evidenced by the extremely small scatter
of the FP solutions displayed by the end products of hierarchical
merger simulations discussed by \cite{CdCC95}. This would suggest that
indeed the fine-tuning mechanism is related to the hierarchical
scenario of formation of galaxies. Alternatively, an explanation for
the FP, avoiding fine-tuning of any type, is given by our model which
includes a small curvature in the FP correlation. However, giving the
clustering scale represented by elliptical galaxies, the scatter of the
FP should be known with much higher accuracy.

\acknowledgments

We would like to thank Drs. J. Dubinski and G. Blumenthal for fruitful
discussions on the subject of this paper. C.C.D. also thanks F.L. de
Sousa.  C.C.D. and A.L.B.R. acknowledge fellowships from FAPESP under
grants 96/03052-4 and 97/13277-6, respectively. This work was partially
supported by CNPq and PRONEX-246.

\figcaption{Projection in the $\kappa$-space of the data presented by
BBFN97. The symbols are as follow: open circle - groups dominated by
elliptical galaxies ; closed circle - elliptical galaxies; open square
- spiral galaxies; closed square - clusters of galaxies; star -
globular clusters; open triangle - groups dominated by spiral
galaxies.  Panel (a) shows $\kappa_{1}\times\kappa_{3}$, where the
dotted lines indicate the variation of $\kappa_{2}$ from -2.5 to 5.0.
For both projections the 2-VT model is constrained by C$^{\star} =
8.28$ and b = 200. Panel (b) displays the projection
$\kappa_{1}\times\kappa_{2}$, where the dotted lines represent
different values of $\kappa_{3}$ as indicated.  }


\begin{thebibliography}{DUM}

\bibitem [Bellazzini 1998]{bel98} Bellazzini, M., 1998, New Astron. 3, 219.
\bibitem [Bender, Burstein \& Faber 1992]{ben92} Bender, R., Burstein, D., \& Faber, S. M., 1992, ApJ, 399, 462.
\bibitem [Binney \& Tremaine 1987]{bet87} Binney, J. \& Tremaine, S., 1987, ``Galactic Dynamics'', Princeton Series in Astrophysics.
\bibitem [Burstein et al. (1997)]{bur97} Burstein, D., Bender, R., Faber, S. M., \& Nolthenius, 1997, AJ, 114, 1365.
\bibitem [Capelato, de Carvalho \& Carlberg (1995, 1997)]{CdCC95} Capelato, H.V., de Carvalho, R.R., Carlberg, R.G., 1995, ApJ, 451, 525.
\bibitem [Capelato, de Carvalho \& Carlberg 1997]{CdCC97} Capelato, H.V., de Carvalho, R.R., Carlberg, R.G., 1997, in ``Galaxy Scaling Relations'', p. 331, L.N. da Costa \& A. Renzini Eds., Springer-Verlag.
\bibitem [Ciotti 1991]{c91} Ciotti,L., 1991, A\&A, 249, 99.
\bibitem [Ciotti, Lanzoni \& Renzini (1996)]{clr96} Ciotti,L., Lanzoni,B., Renzini, A., 1996, MNRAS, 282, 1.
\bibitem [Ciotti \& Lanzoni 1997]{cl96} Ciotti,L., Lanzoni,B., 1997, A\&A, 321, 724.
\bibitem [Djorgovski \& Davis 1987]{djo87} Djorgovski, S. G. \& Davis, M., 1987, ApJ, 313, 59.
\bibitem [Dressler et al. 1987]{dre87} Dressler, A., Lynden-Bell, D., Burstein, D., Davies, R.L., Faber, S., Terlevich, R.J., Wegner, G., 1987, ApJ, 313, 42.
\bibitem [Graham \& Colless 1997]{gc97} Graham, A., Colless, M., 1997, MNRAS, 287,221.
\bibitem [Hjorth \& Madsen 1995]{hm95} Hjorth, J., Madsen, J., 1995, ApJ, 445, 55.
\bibitem [Kritsuk (1997)]{k97} Kritsuk, A.G., 1997, MNRAS, 284, 327.
\bibitem [Limber 1959]{limb59} Limber, D.N., 1959, ApJ, 130, 414.
\bibitem [Pahre, Djorgovski \& de Carvalho 1998]{pah98} Pahre, M. A., Djorgovski, S. G. \& de Carvalho, R. R., 1998, AJ, 116, 1591.
\bibitem [Pryor \& Meylan 1993]{pry93} Pryor, C., \& Meylan, G., 1993, ASP Conf. Ser. Vol. 50, Structure and Dynamics of Globular Clusters, ed. S. Djorgovski \& G. Meylan, p. 357.
\bibitem [Schaeffer et al. 1993]{sch93} Schaeffer, R., Maurogordato, S., Cappi, A., Bernardeau, F., 1993, MNRAS, 263, L21.
\bibitem [Smith 1980]{smith} Smith Jr., H., 1980, ApJ, 241, 63
\bibitem [Spitzer 1969]{spi69} Spitzer,L., 1969, ApJ, 158, L141.
\end{thebibliography}
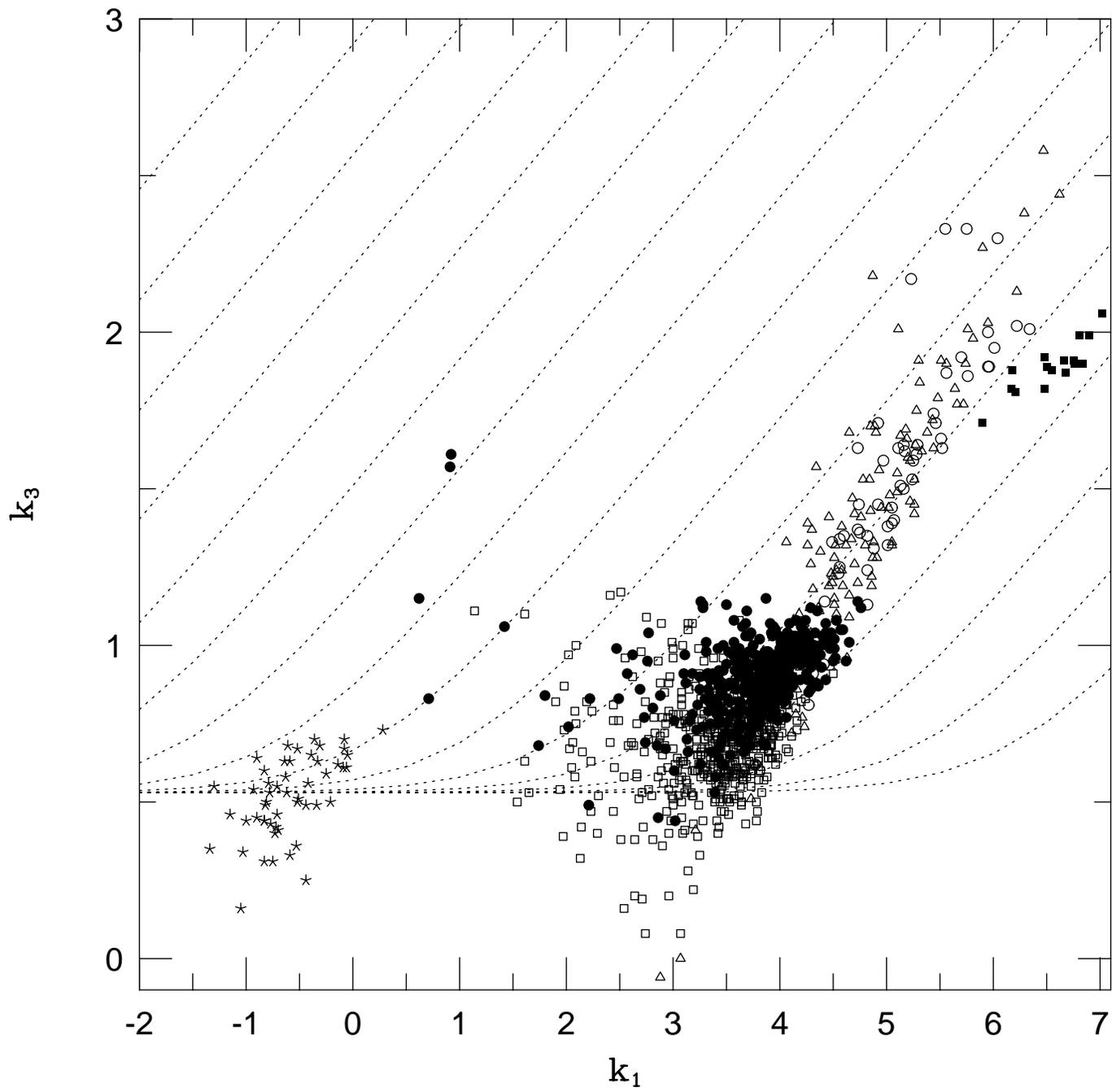
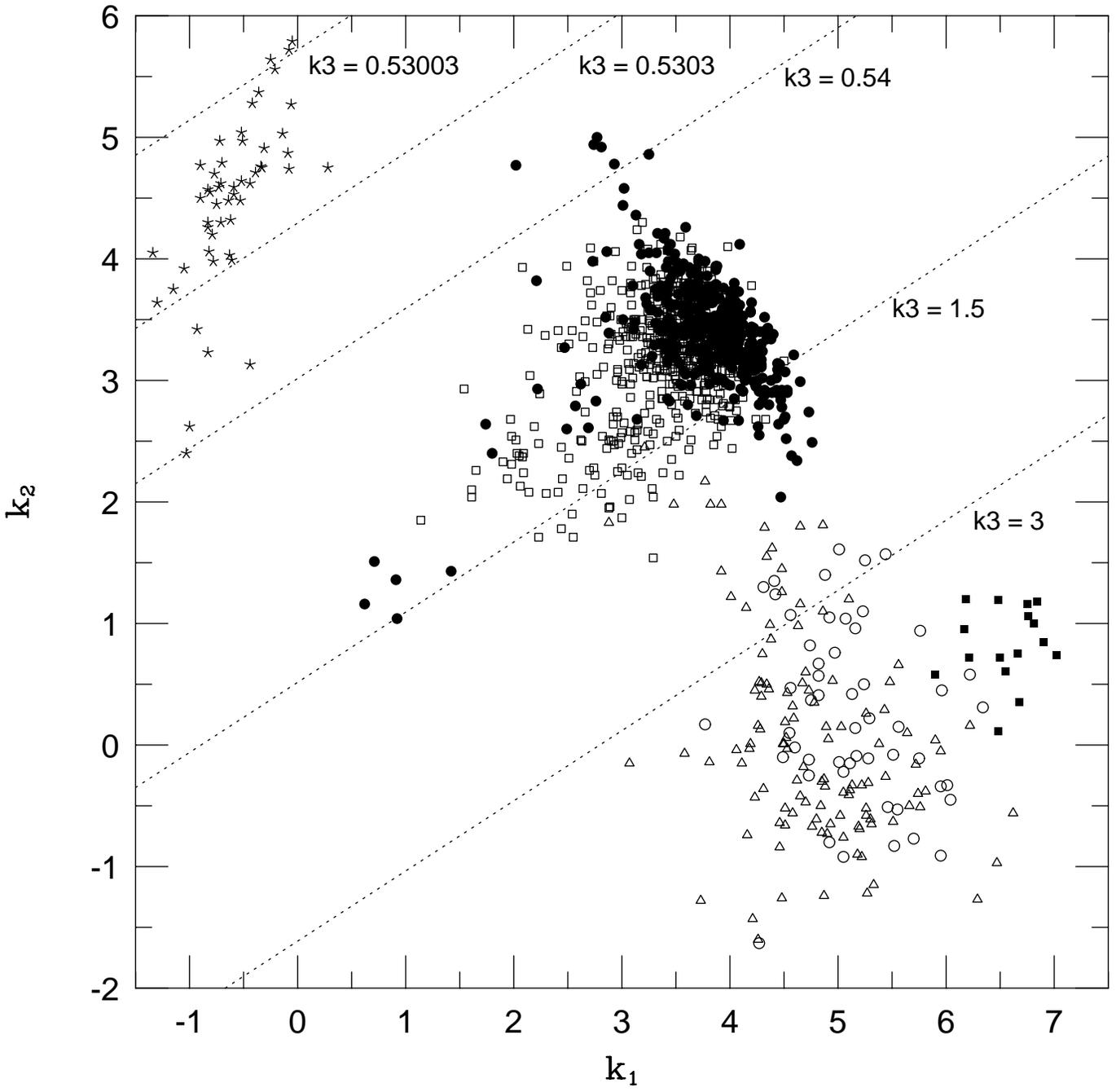
\end{document}